\documentclass[aps,showpacs]{revtex4}
\usepackage{graphicx}
\usepackage{latexsym}
\newcommand{\beq}{\begin{equation}}
\newcommand{\beqn}{\begin{eqnarray}}
\newcommand{\eeq}{\end{equation}}
\newcommand{\eeqn}{\end{eqnarray}}

\def\be{\begin{equation}}
\def\ee{\end{equation}}
\def\bea{\begin{eqnarray}}
\def\eea{\end{eqnarray}}
\begin{document}

\title{Inflaton perturbations in brane-world cosmology with induced gravity} 

\author{Kazuya Koyama$^1$ and Shuntaro Mizuno$^2$}

\address{
 $^1$Institute of Cosmology \& Gravitation,
University of Portsmouth,
Portsmouth~PO1~2EG, United Kingdom\\
$^2$ Research Center for the Early Universe (RESCEU), Graduate
School of
Science, The University of Tokyo, Tokyo 113-0033, Japan}

\begin{abstract}
We study cosmological perturbations in the brane models 
with an induced Einstein-Hilbert term on a brane. 
We consider an inflaton confined to a de Sitter brane in a five-dimensional
Minkowski spacetime.
Inflaton fluctuations excite Kaluza-Klein modes of 
bulk metric perturbations with mass $m^2 = -2(2\ell-1) (\ell +1) H^2$ and 
$m^2 = -2\ell(2\ell+3) H^2$ where $\ell$ is an integer.
There are two branches ($\pm$ branches) of solutions for the 
background spacetime.
In the $+$ branch, which includes the self-accelerating universe,
a resonance appears for a mode with $m^2 = 2 H^2$ due to 
a spin-0 perturbation with $m^2 = 2H^2$.
The self-accelerating universe has a distinct feature because 
there is also a helicity-0 mode of spin-2 perturbations with $m^2 = 2H^2$.
In the $-$ branch, which can be thought as the Randall-Sundrum type
brane-world with the high energy quantum corrections,
there is no resonance. At high energies, 
we analytically confirm that four-dimensional Einstein gravity is recovered, 
which is related to the disappearance of van Dam-Veltman-Zakharov discontinuity 
in de Sitter spacetime. On sufficiently small scales, we confirm that 
the lineariaed gravity on the brane is well described by the Brans-Dicke theory
with $\omega=3Hr_c$ in $-$ branch and $\omega = -3H r_c$ in $+$ branch, 
respectively, which confirms the existence of the ghost in $+$ branch.
We also study large scale perturbations. In $+$ branch, the resonance 
induces a non-trivial anisotropic stress on the brane via the projection 
of Weyl tensor in the bulk, but no instability is shown to exist on the brane.
\end{abstract}



\maketitle

\section{Introduction}

There has been tremendous interest over the last several
years in the brane-world scenario where we are assumed to be 
living on a four-dimensional hypersurface (brane) in a 
higher-dimensional spacetime (bulk) \cite{Roy_review}.
The simplest example of this scenario is proposed
by Randall and Sundrum where there is a brane embedded
in a five-dimensional anti-de Sitter (AdS) spacetime 
\cite{RS99}. They showed that the four-dimensional gravity
is recovered at low energies, because gravity is confined
to a single positive-tension brane even if the extra 
dimension is not compact. Therefore, 
in order to see the deviation from the conventional
four-dimensional gravity, it is necessary to investigate
the phenomena occurred at high energies. 

One of the most promising tools to extract the information
of the extra-dimension is the primordial density
fluctuations generated in the period of inflation
in the early universe. In the inflation model 
where inflation is driven by an inflaton field
confined to the brane, the amplitude of 
the curvature perturbation is calculated in the 
extremely slow-roll limit where the 
coupling between inflaton field fluctuation 
and bulk perturbations can be neglected \cite{MWBH}.
This work is proceeded to go beyond the zero-th order
slow-roll approximation by solving the bulk metric
perturbations classically \cite{KLMW,KMW}. 
It is shown that we could have significant effects 
from the back-reaction due to the coupling to 
five-dimensional gravitational 
perturbations. On the other hand, it was pointed out that 
the localized matter on a brane can induce gravity on the brane 
via quantum loop corrections at high energies \cite{induced_gravity}. 
This induced gravity can act as the ultra-violet cut-off for the 
inflaton perturbations. 

Based on the induced gravity scenario, Dvali, Gabadadze, and Porrati
(DGP) proposed a brane-world model with induced gravity
 \cite{DGP} in which the four-dimensional brane is
embedded in a five-dimensional Minkowski spacetime. 
(For a review of the phenomenology of DGP model, see
\cite{Lue_review}.)
In the presence of the induced gravity,
according to the embedding of the brane in the bulk,
there appear two branches of background solutions. In $+$ branch, the size of the
four-dimensional hypersurface becomes minimum at the brane while, in $-$
branch, it becomes maximum at the brane. Inflation models in this scenario 
were studied in \cite{inducegw, Papan, Zhang}.

In this paper, we study the behavior of 
five-dimensional metric perturbations 
excited by the inflaton perturbations confined to the brane in 
DGP model.
The solutions in $-$ branch are regarded as 
the high energy limit of the Randall-Sundrum model including 
the curvature corrections on a brane, depending 
on the parameters. Then our analysis 
can be used to discuss the generation of primordial fluctuations
in these models. The solutions in $+$ branch are largely used
to explain the present cosmic acceleration
without introducing dark energy in the DGP model
\cite{DGP_FRW}, but it turns out that they are plagued by the ghost instability
\cite{Nicolis_Rattazzi,Luty_Porrati_Rattazzi,Koyama_ghost,Gorbunov_Koyama_Sibiryakov,
Charmousis}.
However, the study of the gravitational property of this model 
in the presence of matter and cosmic expansion is still 
very limited \cite{Lue_Starkman, KM, Kaloper} so it is also worth considering 
the solutions in this branch. 

The structure of the rest of the paper is as follows.
In section II,  we derive the basic equations
for the five-dimensional metric perturbation and the boundary
condition imposed at the brane and then summarize the 
solution for the vacuum brane obtained in \cite{Koyama_ghost}.
In section III, we provide the solutions for curvature perturbations on the brane 
in the presence of the scalar field on the brane. Next we
discuss the appearance and disappearance of 
van Dam-Veltman-Zakharov discontinuity in section IV.
In section V we show the comparison of the behavior of 
perturbations on small scales with the Brans-Dicke theory.
The solutions for perturbations on large scales are discussed in section VI. 
Section VII is devoted to conclusions.
Technical details of the calculation of the $m^2 = 2H^2$
mode and the brief summary of the cosmological perturbation
in four-dimensional Brans-Dicke gravity are presented in the 
Appendix A and B, respectively.

\section{Bulk gravitons with a de Sitter brane}

\subsection{Background cosmology}
We consider a four-dimensional brane-world model with a 
five-dimensional infinite-volume bulk. 
The action is given by 
\begin{eqnarray}
S = \frac{1}{2\kappa^2} \int d^5x \sqrt{-g}
{}^{(5)}R + \int d^4 x \sqrt{-\gamma}
\left[\frac{1}{\kappa^2}K + L_{\rm brane}
+ \frac{1}{2 \kappa_4^2}R\right],
\label{bulkbrane_action}
\end{eqnarray}
where $\kappa^2$ is the fundamental five-dimensional
gravitational constant, $K$ is the trace of extrinsic curvature $K_{AB}$
on the brane and $L_{\rm brane}$ is a Lagrangian for brane-localized matter.
In addition, we consider the last term, an intrinsic curvature term 
on the brane which plays a crucial role in this model.
Here, $\kappa_4^2 = 8 \pi G$ is the four-dimensional
gravitational constant.
We define a crossover scale $r_c$ as
\begin{eqnarray}
r_c = \frac{\kappa^2}{2\kappa_4^2}.
\label{def_rc}
\end{eqnarray}

We are interested in the gravitational property
of this model in a cosmological setting.
By assuming the flat Friedmann-Robertson-Walker
metric on the brane and neglecting the effect
from the five-dimensional Weyl tensor
in the background,
the following Friedmann like equation
is derived \cite{DGP_FRW}; 
\begin{eqnarray}
H^2 -\frac{\epsilon}{r_c} H = 
\frac{\kappa_4^2} {3} \rho,
\label{effective_Friedman}
\end{eqnarray}
where $H$ is a Hubble parameter on the brane
and $\epsilon=\pm 1$, which is the parameter 
related to the embedding of the brane in the bulk.
We call the case with $\epsilon = 1$ $+$ branch,
while $\epsilon = -1$ $-$ branch.

In this paper, we restrict ourselves to de Sitter background
$H =$ constant.
In this case, the energy density is constant and given by 
\begin{eqnarray}
\rho = \frac{3 H}{\kappa_4^2 r_c}(r_c H -\epsilon).
\end{eqnarray}

The 5D solution for the metric with the four-dimensional de Sitter brane can be obtained
as
\begin{eqnarray}
ds^2 = dy^2 + N(y)^2 \gamma_{\mu\nu} dx^\mu dx^\nu,\;\;\;\;
N(y) = 1 + \epsilon H y,
\label{5dmetric}
\end{eqnarray}
where $\gamma_{\mu\nu}$ is the metric for the de Sitter spacetime and 
the brane is located at $y=0$.
During a slow-roll inflation, the Hubble scale is varying slowly, 
thus we can approximate the background geometry by 
(\ref{5dmetric}).

\subsection{Master variable for perturbations in the Minkowski bulk}
Now let us consider the scalar perturbation. 
In order to solve the perturbations in this background,
it is convenient to use five-dimensional
longitudinal gauge, given by
\begin{eqnarray}
ds^2 = (1+2A_{yy})dy^2 + 2N(y)A_y dy dt + N(y)^2
\Big[-(1+2A) dt^2 + a(t)^2 (1+2 {\cal{R}}) \delta_{ij}dx^i dx^j\Big],
\end{eqnarray}
where $a (t) =\exp (Ht)$.

In the absence of bulk matter perturbations, 
five-dimensional perturbed Einstein 
equations ${}^{(5)}\delta G^A_{\;\;B} = 0$ are solved in a Minkowski
background if the metric perturbations are derived
from a `master variable', $\Omega$ \cite{Mukohyama_variable} (see also \cite{Kodama});
\begin{eqnarray}
A &=& -\frac{1}{6 a N} \left(2\Omega'' - \frac{N'}{N} \Omega'
+ \frac{1}{N^2} \ddot{\Omega}\right),\nonumber\\
A_y &=& \frac{1}{a N^2} 
\left(\dot{\Omega}'- \frac{N'}{N} \dot{\Omega}\right),\nonumber\\
A_{yy} &=& \frac{1}{6 a N} 
\left(\Omega'' - 2\frac{N'}{N}\Omega ' + \frac{2}{N^2} \ddot{\Omega} \right),
\nonumber\\
{\cal{R}} &=& \frac{1}{6a N} 
\left(\Omega'' + \frac{N'}{N} \Omega ' - \frac{1}{N^2}
\ddot{\Omega}\right),
\label{mitric_pert_in_5dlt}
\end{eqnarray}
where dot and prime denote derivatives with respect to $t$ and $y$,
respectively.
The perturbed five-dimensional Einstein equations
yield a single wave 
equation governing the evolution of the master variable
$\Omega$ in the bulk:
\begin{eqnarray}
\ddot{\Omega} - 3H \dot{\Omega}-
\left(\Omega'' - 2 \frac{N'}{N}\Omega'\right)
+\frac{k^2}{a^2}\Omega =0.
\end{eqnarray}
Solutions for the master equation can be separated
into eigenmodes of the time-dependent equation on the
brane and bulk mode equation:
\begin{eqnarray}
\Omega(t,y;\vec{x}) = \int d^3 \vec{k} dm g_{m} (t) f_m (y) e^{ikx},
\end{eqnarray}
where
\begin{eqnarray}
\ddot{g}_{m} &-& 3H\dot{g}_{m} +\left[m^2 +\frac{k^2}{a^2}
 \right]g_m =0,\\
f_m'' &-& 2\frac{N'}{N} f'_m + \frac{m^2}{N^2} f_m =0.
\end{eqnarray}

\subsection{Boundary condition for $\Omega$}
In order to solve $\Omega$, we must specify the boundary
condition for $\Omega$ from the junction conditions
at the brane. The junction conditions in this model 
have been found already in literatures 
(see \cite{Deffayet_perturbation, Deffayet_vDVZdis_FRW}), 
but, we re-derive these so that the property of linearized gravity 
is transparent.

For simplicity, we consider the case where the
anisotropic stress of the matter 
on the brane does not exist. There are two 
important contributions to the junction conditions.
One is that, from Eq.~(\ref{def_tilde_t_munu}), we must take into 
account not only the matter energy momentum, but also the
contribution from the induced gravity on the brane.
The other is that we should take into account the  
brane bending $\xi$ \cite{GT,Tanaka_DGP}. 
The metric perturbations in five-dimensional longitudinal 
gauge evaluated at $y=0$ is {\it not} the induced metric 
perturbations on the brane. This is due to the anisotropic 
stress coming from the induced gravity.
The junction conditions are given by
\begin{eqnarray}
&&\frac{1}{a} \left[\ddot{{\cal{F}}} + 2H 
\dot{{\cal{F}}}\right] - 2 \ddot{\xi} - 4H \dot{\xi}
+ 6H^2 \xi - \frac{4}{3}\frac{k^2}{a^2}\xi
= \kappa^2(\delta p_{\rm m}+ \delta p_{\rm g}),
\label{junc_1_ii}\\
&&-\frac{1}{a}\dot{{\cal{F}}} + 2 \dot{\xi}
-2H \xi = \kappa^2 (\delta q_{\rm m}+ \delta q_{\rm g}),
\label{junc_1_0i}\\
&&-\frac{1}{a}\left[3H \dot{{\cal{F}}} +
\frac{k^2}{a^2}{\cal{F}} \right]
 + 6H \dot{\xi} -6H^2 \xi +2 \frac{k^2}{a^2}\xi
= \kappa^2 (\delta \rho_{\rm m}+ \delta \rho_{\rm g})
\label{junc_1_00},
\end{eqnarray}
where
\begin{equation}
{\cal{F}} = \Omega' - \epsilon H \Omega.
\label{def_cal_F} 
\end{equation}
Here the brane bending mode is given by
\begin{equation}
\xi = -r_c (A_{(b)} +{\cal{R}}_{(b)} ),
\label{def_xi}
\end{equation}
where
$A_{(b)}$ and ${\cal{R}}_{(b)}$ are the induced metric 
perturbations on the brane that are related with the metric 
perturbations in five-dimensional longitudinal gauge as
\begin{eqnarray}
A_{(b)} &=& \frac{1}{1 -2 \epsilon H r_c} \Big[     
(1- \epsilon H r_c) A + \epsilon H r_c {\cal{R}} \Big], 
\label{A_b_ito_AR}\\
{\cal{R}}_{(b)} &=& \frac{1}{1 -2 \epsilon H r_c} \Big[     
(1- \epsilon H r_c) {\cal{R}} + \epsilon H r_c A \Big],
\label{R_b-ito_AR}
\end{eqnarray}
where ${\cal{R}}$ and $A$ are metric perturbations 
in five-dimensional longitudinal gauge 
evaluated at $y=0$, which can be expressed by $\Omega$ as 
\begin{eqnarray}
{\cal{R}} &=& \frac{1}{6 a} \Big[   
3 \epsilon H {\cal F} - 3H (\dot{\Omega} -H \Omega) +\frac{k^2}{a^2} \Omega
\Big], 
\label{R-ito_Fomega}\\
A &=& \frac{1}{6 a} \Big[  
-3 \epsilon H {\cal F} -3 \ddot{\Omega} + 6H \dot{\Omega} -3 H^2 \Omega
-2 \frac{k^2}{a^2} \Omega
\Big].
\label{A-ito_Fomega}
\end{eqnarray}
$\delta \rho_m, \delta p_m$ and $\delta q_m$
are the components of the perturbed energy momentum 
tensor of the matter on the brane and 
\begin{eqnarray}
&&\delta p_{\rm g} = \frac{2}{\kappa_4^2}
\left[-3H^2 A_{(b)} -H \dot{A}_{(b)} + 
\ddot{{\cal{R}}}_{(b)} + 3H\dot{{\cal{R}}}_{(b)}
+\frac{k^2}{3a^2}
( A_{(b)} + {\cal{R}}_{(b)} )\right],
\label{def_rho_g}\\
&&\delta q_{\rm g} = \frac{2}{\kappa_4^2}
(H A_{(b)} - \dot{{\cal{R}}}_{(b)} ),
\label{def_q_g}\\
&& \delta \rho_{\rm g} = \frac{2}{\kappa_4^2}
\left[3H(- \dot{{\cal{R}}}_{(b)} + H A_{(b)})
- \frac{k^2}{a^2} {\cal{R}}_{(b)}\right],
\end{eqnarray}
are contributions from the induced gravity.
If we do not take into account
the induce gravity term which corresponds to
the limit $r_c  \to \infty$, the quantities
$\xi, \delta p_{\rm g}, \delta q_{\rm g}$ and $\delta \rho_{\rm g}$
can be neglected and the junction conditions
in Randall-Sundrum type brane-world 
\cite{Deffayet_perturbation,Koyama_late_time,KLMW}
are reproduced.

Defining ${\cal{G}}$ as
\begin{eqnarray}
{\cal{G}} = (1-2\epsilon H r_c)
{\cal{F}}-r_c (2 H^2 -m^2) \Omega,
\label{def_cal_G}
\end{eqnarray}
the junction conditions 
(\ref{junc_1_ii}),(\ref{junc_1_0i}),(\ref{junc_1_00})
are amazingly simplified as
\begin{eqnarray} 
&&-3H \dot{{\cal{G}}} - \frac{k^2}{a^2}{\cal{G}}
= \kappa^2 a \delta \rho_{\rm m}
,\label{junc_2_gen_00}\\
&&-\dot{{\cal{G}}} = \kappa^2 a \delta q_{\rm m}
,\label{junc_2_gen_0i}\\
&&\ddot{{\cal{G}}} + 2H \dot{{\cal{G}}} =
\kappa^2 a \delta p_{\rm m}.
\label{junc_2_gen_ii}
\end{eqnarray}

\subsection{Solutions for a vacuum brane}
Before discussing the case with the matter perturbations
on the brane, we briefly summarize the case 
for a vacuum brane, $\delta T^{\mu}_{\;\;\nu} = 0$
considered in \cite{Koyama_ghost,Gorbunov_Koyama_Sibiryakov}.
We will only consider a positive tension brane which means 
$Hr_c >1$ for $+$ branch and $Hr_c >0$ for $-$ branch.
The solutions obtained here will serve as homogeneous 
solutions when we include matter perturbations.

For the vacuum brane, the boundary conditions for $\Omega$
given by Eqs.~(\ref{junc_2_gen_00}), (\ref{junc_2_gen_0i})
and (\ref{junc_2_gen_ii}) reduce to a single
boundary condition on the master variable,
\begin{eqnarray} 
{\cal{G}} = 0.
\label{junc_G}
\end{eqnarray} 
For both branches, we find a tower of continuous massive modes
starting from $m^2 = (9/4) H^2$, which are the Kaluza-Klein
(KK) modes of the spin-2 perturbations.

In $+$ branch ($\epsilon = 1$), in addition to continuous modes, 
there are two discrete modes which are normalizable.
One is the mode with $m^2 = 2 H^2$ which is the spin-0
perturbation, and the other is the mode with 
$m_d^2 = H^2 (3Hr_c-1) (Hr_c)^{-2}$ which is the 
helicity-0 excitation of the spin-2 perturbations \cite{KK}.
For $Hr_c > 1$, the mass $m_d^2$ is in the range
$0 < m_d^2 < 2 H^2$. If $Hr_c = 1$, the helicity-0 
excitation of the spin-2 perturbation has mass
$m_d^2 = 2H^2$, which coincides with the mass of the
spin-0 perturbation. Then, there is a resonance 
between these two modes. It was also shown that the spin-2
helicity-0 mode becomes a ghost for $Hr_c >1$. In Figure 1, 
we summarize the mass spectrum for $1 \leq Hr_c $ in $+$ branch.

\begin{figure}[h]
\centerline{
\includegraphics[width=6cm]{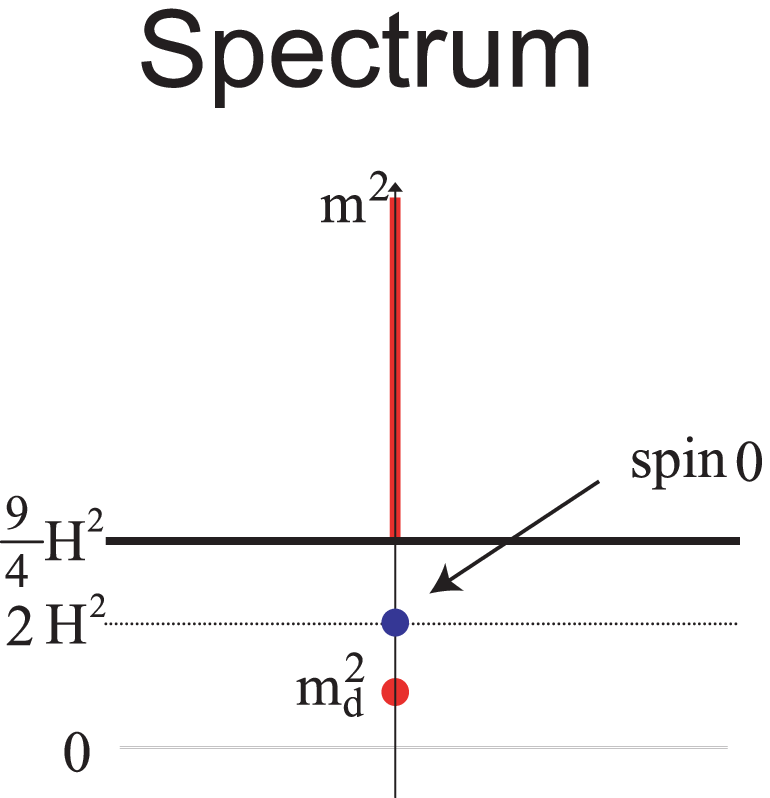}}
\caption{Summary of the mass spectrum of the
scalar perturbations in $+$ branch \cite{Koyama_ghost}. 
Spin-2 perturbation has continuous modes with $m^2 \geq (9/4)H^2$
and a discrete mode with $m^2 = m_d^2$ while spin-0 perturbation
has $m^2 = 2H^2$.
In the limit
$Hr_c \to 1$, both the helicity-0 excitation
of spin-2 perturbation and the spin-0 perturbation have mass
$m^2 = 2H^2$ and there is a resonance 
\cite{Gorbunov_Koyama_Sibiryakov}.
}
\label{fig1}
\end{figure}

\section{Scalar field on the brane}

\subsection{Bulk scalar modes}
Until now, we have not specified the matter on the brane.
In the following, we model the 
matter on the brane as a canonical scalar field $\phi$
with potential $V(\phi)$ whose homogeneous part gives
approximately de Sitter universe on the brane
\cite{KLMW,KMW}.

In this case, the junction conditions 
(\ref{junc_1_ii}), (\ref{junc_1_0i}), (\ref{junc_1_00})
reduce to 
\begin{eqnarray} 
&&-3H \dot{{\cal{G}}} - \frac{k^2}{a^2}{\cal{G}}
= \kappa^2 a [\dot{\phi} \dot{\delta \phi} + {V,}_\phi \delta \phi]
,\label{junc_2_00}\\
&&\dot{{\cal{G}}} = \kappa^2 a \dot{\phi} \delta \phi
,\label{junc_2_0i}\\
&&\ddot{{\cal{G}}} + 2H \dot{{\cal{G}}} =
\kappa^2 a [\dot{\phi} \dot{\delta \phi} - {V,}_\phi  \delta \phi].
\label{junc_2_ii}
\end{eqnarray}

These equations can be thought as the boundary conditions
for $\Omega$. Combining the junction conditions
(\ref{junc_2_00}), (\ref{junc_2_0i}), (\ref{junc_2_ii}),
we get an evolution equation for ${\cal{G}}$:
\begin{eqnarray}
\ddot{{\cal{G}}} -
\left(H + 2\frac{\ddot{\phi}}{\dot{\phi}}\right) \dot{{\cal{G}}}
+ \frac{k^2}{a^2} {\cal{G}}  =0.
\label{evol_eq}
\end{eqnarray}

This gives the boundary condition for the time dependence 
of the master variable $\Omega$. 
Assuming that $\phi$ is slow-rolling, so that 
$|\ddot{\phi}/\dot{\phi}| \ll H$ in Eq.~(\ref{evol_eq}),
which is valid for the de Sitter universe,
the solution for ${\cal{G}}$ is
\begin{eqnarray}
{\cal{G}} = C_1 \frac{\cos (-k\eta)}{-k\eta}
+ C_2 \frac{\sin (-k\eta)}{-k\eta}.
\label{g_junc_cond}
\end{eqnarray}
We use the formulae for summation of Bessel functions, 
\begin{eqnarray}
\sum_{\ell=0}^{\infty} (-1)^\ell \left(2\ell + \frac{3}{2}\right)
z^{- \frac{3}{2}} J_{2\ell + \frac{3}{2}} (z) &=& 
\sqrt{\frac{1}{2\pi}} \frac{\sin z}{z}, \nonumber\\
\sum_{l=0}^{\infty} (-1)^\ell \left(2\ell + \frac{1}{2}\right)
z^{- \frac{3}{2}} J_{2\ell +  \frac{1}{2}} (z) &=& 
\sqrt{\frac{1}{2\pi}} \frac{\cos z}{z}.
\label{tri_formula}
\end{eqnarray}
These show that an infinite sum of mode functions
\begin{eqnarray}
g_m = (-k \eta)^{-3/2} J_{\nu} (- k \eta),\;\;\;\;{\rm where}
\;\; \nu^2 = \frac{9}{4}- \frac{m^2}{H^2},
\end{eqnarray}
can satisfy the boundary condition
imposed on ${\cal{G}}$, where the spectrum of KK modes
is given by
\begin{eqnarray}
\frac{m^2}{H^2} &=& -2(2\ell-1)(\ell+1)\;\;\;\;\;{\rm for}\;\;\; C_1,
\nonumber\\
\frac{m^2}{H^2} &=& -2\ell(2\ell+3)\;\;\;\;\;{\rm for}\;\;\; C_2.
\end{eqnarray}

Figure~2 shows the mass spectrum of the scalar 
perturbation supported by the scalar field on the
brane. Since there is a mode with $m^2 = 2H^2$
regardless of the value of $Hr_c$ in the $+$ branch,
the resonance inevitably appears. 
This gives qualitative difference
between $-$ branch and $+$ branch. In addition, 
for $Hr_c =1$, that is, on the self-accelerating 
universe, the spin-0 and spin-2 modes are 
already degenerates. Thus the behavior of gravity 
is again qualitatively different from $Hr_c >1$. 

\begin{figure}[h]
\centerline{
\includegraphics[width=12cm]{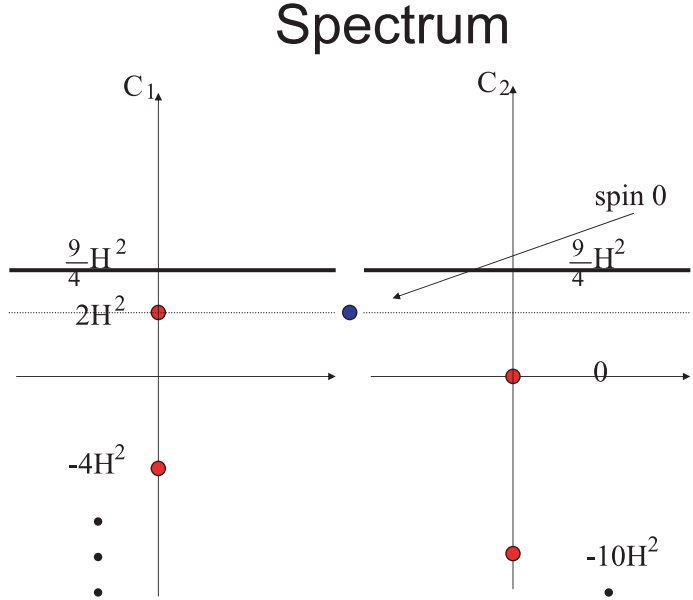}}
\caption{Summary of the mass spectrum of the
scalar perturbation sourced by the scalar
field on the brane. For the modes $C_1$,
there is a mode with $m^2 = 2H^2$ regardless
of the value of $Hr_c$. Since there is another
mode with $m^2 = 2H^2$ by the spin-0 perturbation
in the $+$ branch, there is a resonance. In addition,
for $Hr_c=1$, the helicity-0 mode of spin-2 perturbations
also has $m^2=2H^2$.
}
\label{fig2}
\end{figure}
In order to examine the behavior of the master variable $\Omega$,
we will consider $-$ branch ($\epsilon = -1$)
and $+$ branch ($\epsilon = 1$)
separately in the following.

\vspace{3mm}
{\bf $-$ branch ($\epsilon = -1$)}
\vspace{2mm}

Since there is a horizon at $y=1/H$ in this branch, we should choose the 
solution in the $y$-direction so that the metric perturbations remain small 
as $y \to 1/H$.
By imposing the normalizability condition on $f_m$,
\begin{eqnarray}
\int_0 ^{1/H} N(y)^{-4} |f_m (y)|^2 dy < \infty,\;\;\;\;\;\;\;\;\;\;
{\rm with} \;\;\;\;\;N(y) = 1-Hy,
\end{eqnarray}
we obtain the solution for $\Omega$ in the bulk
(for $y \geq 0$) as 
\begin{eqnarray}
\Omega (\eta, y) =
&-&C_1 \frac{\sqrt{2\pi}}{H} \sum_{\ell = 0} ^{\infty}
(-1)^{\ell} \left(2\ell + \frac{1}{2}\right)
\frac{(1-Hy)^{2+2\ell}}{(2 \ell+1) \{2( \ell +1)H r_c +1\}} 
(- k \eta)^{- \frac{3}{2}} J_{2 \ell +  \frac{1}{2}} (-k\eta)\nonumber\\
&-&C_2\frac{\sqrt{2 \pi}}{H} \sum_{\ell = 0} ^{\infty}
(-1)^{\ell} \left(2\ell + \frac{3}{2}\right)
\frac{(1-Hy)^{3+2\ell}}{2( \ell + 1) \{(2\ell+3)H r_c +1\}} 
(- k \eta)^{- \frac{3}{2}} J_{2 \ell +  \frac{3}{2}} (-k\eta).
\label{omega_minus}
\end{eqnarray}

\vspace{3mm}
{\bf $+$ branch ($\epsilon = 1$)}
\vspace{2mm}

In $+$ branch, the bulk is infinite. 
We should choose the solution in the $y$-direction so that the 
metric perturbations remain small as $y \to \infty$ and the mode
is normalizable for a single brane. 
We impose the normalizability condition on $f_m$ as 
\begin{eqnarray}
\int_0 ^{\infty} N(y)^{-4} |f_m (y)|^2 dy < \infty,\;\;\;\;\;\;\;\;\;\;
{\rm with} \;\;\;\;\;N(y) = 1+Hy.
\end{eqnarray}
Because of the existence of the resonance,
we must treat the mode with $m^2 = 2H^2$ separately.
We also treat $Hr_c=1$ case separately as the spin-2 and 
spin-0 degenerate (see Appendix A for derivations).

We obtain the solution for $\Omega$ in the bulk (for $y \geq$ 0) as
\begin{eqnarray}
\Omega (\eta, y) = && \Omega_{(m^2 = 2H^2)} (\eta, y)\nonumber\\
&-&C_1\frac{ \sqrt{2 \pi}}{H} \sum_{\ell = 1} ^{\infty}
(-1)^{\ell} \left(2\ell + \frac{1}{2}\right)
\frac{(1+Hy)^{1-2\ell}}{2 \ell  \{(2 \ell -1)H r_c +1\}} 
(- k \eta)^{- \frac{3}{2}} J_{2 \ell +  \frac{1}{2}} (-k\eta)\nonumber\\
&-&C_2 \frac{ \sqrt{2 \pi}}{H} \sum_{\ell = 0} ^{\infty}
(-1)^{\ell} \left(2\ell + \frac{3}{2}\right)
\frac{(1+Hy)^{2\ell}}{(2 \ell + 1)  (2 \ell H r_c +1)} 
(- k \eta)^{- \frac{3}{2}} J_{2 \ell +  \frac{3}{2}} (-k\eta),\nonumber\\
\label{omega_plus}
\end{eqnarray}
where 
\begin{eqnarray}
\Omega_{(m^2 = 2H^2)}&=& C_1 \frac{(1+Hy)}{2H(Hr_c -1)}
(-k\eta)^{-2}\{\alpha(-k\eta) \sin(-k\eta)
+ \beta(-k\eta)\cos(-k\eta) \}\nonumber\\
&&-C_1 
\frac{(1+Hy)\ln(1+Hy)}{H (Hr_c -1)}
(-k\eta)^{-2} \sin (-k\eta),
\label{omega_plus_m^2_2H^2}\\
\label{alpha_def_1}
\alpha(-k\eta) &=& \int^{-k\eta} d(-k\bar{\eta}) (-k\bar{\eta})^{-2}
\sin (-2k\bar{\eta}),\\
\label{beta_def_1}
\beta(-k\eta) &=& 
\int^{-k\eta} d(-k\bar{\eta}) (-k\bar{\eta})^{-2}
\{\cos (-2k\bar{\eta})-1\},
\end{eqnarray}
for $Hr_c \neq 1$ and 
\begin{eqnarray}
\label{omega_plus_m^2_2H^2_1}
\Omega_{(m^2 = 2H^2)}&=&
 -\frac{C_1}{2H} (1+Hy)(-k\eta)^{-2}
\{\tilde{\alpha}(-k\eta) \sin(-k\eta) 
+ \tilde{\beta}(-k\eta) \cos(-k\eta)\}
\nonumber\\
&&+\frac{C_1}{2H} (1+Hy) \ln (1+Hy)(-k\eta)^{-2}
\{\alpha(-k\eta) \sin(-k\eta) + \beta(-k\eta)
\cos (-k\eta)\}
\nonumber\\
&&-\frac{C_1}{2H} 
(1+Hy)\{\ln (1+Hy)\}^2 (-k\eta)^{-2}
 \sin(-k\eta),\\
\label{alpha_def_2}
\alpha(-k\eta) &=& \int^{-k\eta} d(-k\bar{\eta}) (-k\bar{\eta})^{-2}
\sin (-2k\bar{\eta}),\\
\label{beta_def_2}
\beta(-k\eta) &=& 
\int^{-k\eta} d(-k\bar{\eta}) (-k\bar{\eta})^{-2}
\{\cos (-2k\bar{\eta})-1\},\\
\label{til_alpha_def_2}
\tilde{\alpha}(-k\eta) &=& \alpha(-k\eta)+\frac{1}{2}
\int^{-k\eta} d(-k\bar{\eta}) (-k\bar{\eta})^{-2}
[\alpha(-k\bar{\eta}) \sin(-2k\bar{\eta})
+\beta (-k\bar{\eta}) \{\cos(-2k\bar{\eta})+1\}],\\
\label{til_beta_def_2}
\tilde{\beta}(-k\eta) &=& \beta(-k\eta) +\frac{1}{2}
\int^{-k\eta} d(-k\bar{\eta}) (-k\bar{\eta})^{-2}
[\alpha(-k\bar{\eta}) \{\cos(-2k\bar{\eta})-1\}
-\beta (-k\bar{\eta}) \sin(-2k\bar{\eta}) ],
\end{eqnarray}
for $Hr_c =1$.

\subsection{Curvature perturbation on the brane (general solutions)}
Since we have obtained the general solutions for $\Omega$, 
from Eqs.~(\ref{A_b_ito_AR}), (\ref{R_b-ito_AR}),
(\ref{R-ito_Fomega}) and (\ref{A-ito_Fomega}) the induced 
metric perturbations on the brane can be obtained.
We concentrate on the behavior of the curvature
perturbation on the brane ${\cal{R}}_{(b)}$.
As in the previous section, 
we will consider $-$ branch ($\epsilon = -1$)
and $+$ branch ($\epsilon = 1$) separately.

\vspace{3mm}
{\bf $-$ branch ($\epsilon = -1$)}
\vspace{2mm}

In $-$ branch ($\epsilon = -1$), the curvature perturbation
on the brane is given as
\begin{eqnarray}
{\cal{R}}_{(b)} &=&- \frac{\sqrt{2 \pi}}{6} \frac{C_1}{k}  
\frac{H^2}{(2 H r_c +1)} 
\sum_{\ell = 0} ^{\infty} 
\frac{(-1)^{\ell} (2\ell +\frac{1}{2})}{(2\ell +1)\{2(\ell +1)H
r_c +1\}}
\nonumber\\
&&\times 
[6(\ell +1)(2\ell+1)H r_c (-k\eta)^{-\frac{1}{2}}
J_{2\ell + \frac{1}{2}}(-k\eta) + 
2 (2\ell + 1) (-k\eta)^{\frac{1}{2}} J_{2\ell - \frac{1}{2}} (-k\eta)
-(-k\eta)^{\frac{3}{2}}J_{2\ell - \frac{3}{2}} (-k\eta)]\nonumber\\
&&-\frac{\sqrt{2 \pi}}{6} \frac{C_2}{k}  
\frac{H^2}{(2 H r_c +1)} 
\sum_{\ell = 0} ^{\infty} 
\frac{(-1)^{\ell} (2 \ell + \frac{3}{2})}{2(\ell +1)\{(2\ell +3)H r_c +1\}}
\nonumber\\
&&\times 
[ 6(\ell +1)(2\ell +3)H r_c (-k \eta)^{-\frac{1}{2}} J_{2\ell +\frac{3}{2}}(-k \eta)
+2 (2\ell +2) (-k \eta)^{\frac{1}{2}} J_{2\ell + \frac{1}{2}} (-k \eta)
- (-k \eta)^{\frac{3}{2}} J_{2\ell - \frac{1}{2}} (-k \eta)].
\nonumber\\
\label{R_brane_minus}
\end{eqnarray}

\vspace{3mm}
{\bf $+$ branch ($\epsilon = 1$)}
\vspace{2mm}

In  $+$ branch ($\epsilon = 1$), 
the curvature perturbation on the brane is given as
\begin{eqnarray}
{\cal{R}}_{(b)} &=&{\cal{R}}_{(b)(m^2 = 2H^2)}\nonumber\\
&&- \frac{\sqrt{2 \pi}}{6} \frac{C_1}{k}  
\frac{H^2}{(2 Hr_c -1)} 
\sum_{\ell = 1} ^{\infty} 
\frac{(-1)^{\ell} (2\ell + \frac{1}{2})}{2\ell\{(2\ell -1)H
r_c +1\}}
\nonumber\\
&&\times 
[\{ 6\ell (2\ell -1) H r_c +12\ell+3 \}(-k\eta)^{- \frac{1}{2}}
J_{2\ell +  \frac{1}{2}}(-k\eta) -
2 (2\ell + 1) (-k\eta)^{ \frac{1}{2}} J_{2\ell -  \frac{1}{2}} (-k\eta)
+(-k\eta)^{ \frac{3}{2}}J_{2\ell - \frac{3}{2}} (-k\eta)]\nonumber\\
&&-\frac{\sqrt{2 \pi}}{6} \frac{C_2}{k}  
\frac{H^2}{(2 H r_c  -1)} 
\sum_{\ell = 0} ^{\infty} 
\frac{(-1)^{\ell} (2 \ell + \frac{3}{2})}{(2\ell +1)\{2\ell H r_c +1\}}
\nonumber\\
&&\times 
[ \{6\ell (2\ell + 1)H r_c +12\ell +9\} 
(-k \eta)^{-\frac{1}{2}} J_{2\ell + \frac{3}{2}}(-k \eta)
-4 (\ell +1) (-k \eta)^{\frac{1}{2}} J_{2\ell + \frac{1}{2}} (-k \eta)
+(-k \eta)^{\frac{3}{2}} J_{2\ell - \frac{1}{2}} (-k \eta)],\nonumber\\
\label{R_brane_plus}
\end{eqnarray}
where
\begin{eqnarray}
{\cal{R}}_{(b)(m^2 = 2H^2)} &=& -\frac{C_1}{6k} \frac{H^2}{(2 H r_c-1)(H
 r_c-1)}
\bigl[3(H r_c-1)(-k\eta)^{-1}
\sin (-k \eta) \nonumber\\
&&+\frac{\alpha(-k \eta)}{2} \{-3(-k\eta)^{-1} \sin (-k \eta) +
3 \cos (-k \eta) + (-k \eta) \sin (-k \eta)\}
\nonumber\\
&&
+\frac{\beta(-k \eta)}{2}\{-3(-k\eta)^{-1} \cos(-k\eta)
-3 \sin(-k\eta) +(-k\eta) \cos (-k\eta)\}\bigr],
\label{rb_m2_2h2}
\end{eqnarray}
for $H r_c \neq 1$ and 
\begin{eqnarray}
{\cal{R}}_{(b)(m^2 = 2H^2)} &=&
\frac{C_1 H^2}{12k}\bigl[
-6(-k\eta)^{-1} \sin(-k\eta) \nonumber\\
&&+\tilde{\alpha}(-k\eta) \{-3 (-k\eta)^{-1}\sin(-k\eta)
+ 3 \cos(-k\eta) + (-k\eta) \sin(-k\eta)\}
\nonumber\\
&&+\tilde{\beta}(-k\eta) 
\{-3(-k\eta)^{-1}\cos(-k\eta) -3 \sin(-k\eta)
+(-k\eta)\cos(-k\eta)\}\bigr],
\label{rb_m2_2h2_hrc1}
\end{eqnarray}
for $H r_c =  1$.  

We should note that there are homogeneous solutions that 
satisfy ${\cal G}=0$. These homogeneous solutions also induce 
curvature perturbations on the brane.

\section{van Dam-Veltman-Zakharov discontinuity}
\subsection{Curvature perturbation on the brane in high energy limit}
Since the background cosmology given by  
Eq.~(\ref{effective_Friedman}) recovers the conventional
four-dimensional cosmology based on the general relativity at high energies,
we expect that the gravity behaves like
four-dimensional general relativity in this limit.
Here, we show this fact analytically using our solutions.
If we take the limit $H r_c \to \infty$ for  fixed $-k\eta$,
by using the formula 
Eq.~(\ref{tri_formula}), 
the sum of the infinite ladder of the discrete modes can be performed 
and we obtain
\begin{eqnarray}
{\cal{R}}_{(b)} = -A_{(b)} = -\frac{H}{4 k r_c}
\left(C_1 \cos (-k\eta) + C_2 \sin (-k\eta) \right),
\label{rb_hihg_energy}
\end{eqnarray}
in both branches, which agree with the solutions 
in general relativity Eqs.~(\ref{curv_pert_4DGR_dS}), (\ref{lapse_func_4DGR_dS}). 
Therefore, we see that four-dimensional general relativity is recovered in this high energy limit.
This confirms the result first obtained in Ref.~\cite{Deffayet_vDVZdis_FRW}. 

In this limit, the brane bending vanishes
$\xi =- r_c({A}_{(b)}+ {\cal{R}}_{(b)} ) = 0$. We can 
understand this result in a very simple way as follows.
In terms of $A$ and ${\cal R}$, $A_{(b)}$ and ${\cal R}_{(b)}$ are given by
\begin{equation}
A_{(b)} + {\cal R}_{(b)} = \frac{1}{1-2 \epsilon H r_c} 
(A + {\cal R} ).
\label{bendcurve}
\end{equation}
Then at high energies $Hr_c \gg 1$, 
the brane bending mode is shielded by the curvature of the brane
\cite{Lue_Starkman}.

However, this argument holds only for 
$Hr_c \to \infty$. For a finite $Hr_c$,
if we do {\it not} fix $-k \eta$ and take large 
enough $-k \eta$, we can no longer perform the summation 
of the infinite ladder of the mode and obtain the solutions 
(\ref{rb_hihg_energy}). Thus the theory deviates from general 
relativity. This is related with the well-known 
van Dam-Veltman-Zakharov (vDVZ) discontinuity 
\cite{vanDam:1970vg}
first discussed in the context of the Pauli-Fierz theory 
for a massive spin two field \cite{Pauli_Fierz} and 
also in DGP model \cite{DGP}.

\subsection{van Dam-Veltman-Zakharov discontinuity}
Let us consider small scales limit $-k \eta \gg 1$. 
In this region, $(0,0)$ component of the junction conditions 
Eqs.~(\ref{junc_1_00}) and the solution for the brane bending 
give
\begin{eqnarray}
&&\frac{2}{\kappa_4^2} \frac{k^2}{a^2}{\cal{R}}_{(b)}
= \dot{\phi}\delta \dot{\phi}-\frac{2}{\kappa^2}
\frac{k^2}{a^2}\xi+ \frac{k^2}{a^3}\Omega',
\label{junc_1_00_dvz}\\
&& A_{(b)} + {\cal R}_{(b)}= -\frac{1}{r_c} \xi.
\label{junc_1-ij_dvz}
\end{eqnarray}
On small scales, we can assume 
$H \Omega' \ll (k^2/a^2) \Omega$,
Then, Eq.~(\ref{junc_1_00_dvz}) and (\ref{junc_1-ij_dvz}) becomes completely the same
as the one obtained from Brans-Dicke (BD) theory
(\ref{field_eq_BD_00_s}) and (\ref{field_eq_BD_ij_s})
\cite{KM},
provided that
\begin{equation}
\delta \varphi = (1/\kappa^2) \xi.
\end{equation}
Thus in general we do not recover four-dimensional general relativity
but BD theory. This is due to the fact that four-dimensional gravity
is recovered by a continuum of massive states of five-dimensional 
graviton.

In the high-energy limit, we have essentially shown that the 
BD parameter becomes 
infinity for $Hr_c \to \infty$.  
This results remind us that in the case of massive gravity,
if we introduce a cosmological constant $\Lambda_{(4)}$,
the vDVZ discontinuity
disappears if the limits $m \to 0$ is taken for fixed 
$\Lambda_{(4)}$ \cite{Porrati:2000cp, Higuchi,Kogan:2000uy}.

For a finite $Hr_c$, the theory is not described exactly by general relativity 
and this difference shows up at $-k \eta \to \infty$.
Then we need to know what kind of BD theory is realized for a fixed $Hr_c$.
The value of the BD parameter $\omega$ is obtained only by solving the
equation of the motion for the BD scalar field
(\ref{field_eq_BD_varphi}) because it is the only equation
which includes the BD parameter.
In the BD theory, the BD scalar obeys a simple four-dimensional 
equation of motion. However, in the brane-world model, 
the behavior of the brane bending mode can not be obtained 
unless we solve the five-dimensional perturbations. Now, since 
we have a solution
for $\Omega$, we can find the solutions for the brane
bending mode. In the next section, we identify the BD parameter
using the solution for the brane bending.

\section{Comparison with Brans-Dicke theory}
In this section, we derive the solutions for the brane bending mode 
and identify the BD parameter for the effective theory on small scales. 

\subsection{Brane bending mode}

Here, we show the solutions for the
brane bending mode $\xi$ in $-$ branch and 
$+$ branch separately.

\vspace{3mm}
{\bf $-$ branch ($\epsilon = -1$)}
\vspace{2mm}

In $-$ branch ($\epsilon = -1$), the brane bending $\xi$
is given as

\begin{eqnarray}
\xi  &=& -\frac{\sqrt{2 \pi}}{3} \frac{C_1 H}{k}  
\frac{H r_c}{(2H r_c +1)} 
\sum_{\ell = 0} ^{\infty} 
\frac{(-1)^{\ell} (2\ell +\frac{1}{2})}{(2\ell +1)\{2(\ell +1)H r_c+1\}}
\nonumber\\
&&\times 
[3(\ell+1)(2\ell +1)(-k\eta)^{-\frac{1}{2}}J_{2\ell +\frac{1}{2}}(-k\eta) -2 (2\ell + 1) (-k\eta)^{\frac{1}{2}} J_{2\ell - \frac{1}{2}} (-k\eta)
+(-k\eta)^{\frac{3}{2}}J_{2\ell - \frac{3}{2}} (-k\eta)]\nonumber\\
&&-\frac{\sqrt{2 \pi}}{3} \frac{C_2 H}{k}  
\frac{H r_c}{(2 H r_c +1)} 
\sum_{\ell = 0} ^{\infty} 
\frac{(-1)^{\ell} (2 \ell + \frac{3}{2})}{2(\ell +1)\{(2\ell +3)H r_c +1\}}
\nonumber\\
&&\times 
[3(\ell+1)(2\ell +3)(-k\eta)^{-\frac{1}{2}}J_{2\ell +\frac{3}{2}}(-k\eta) 
-2 (2\ell +2) (-k \eta)^{\frac{1}{2}} J_{2\ell + \frac{1}{2}} (-k \eta)
+ (-k \eta)^{\frac{3}{2}} J_{2\ell - \frac{1}{2}} (-k \eta)].
\label{xi_minus}
\end{eqnarray}

\vspace{3mm}
{\bf $+$ branch ($\epsilon = 1$)}
\vspace{2mm}

In $+$ branch ($\epsilon = 1$), the brane bending $\xi$
is given as
\begin{eqnarray}
\xi &=& \xi_{(m^2 = 2 H^2)} \nonumber\\
&&-\frac{\sqrt{2\pi}}{3}\frac{C_1 H}{k}
\frac{H}{2 H r_c -1} \sum_{\ell = 1}^{\infty}
\frac{(-1)^{\ell} (2 \ell + \frac{1}{2})}{2 \ell\{(2 \ell -1) Hr_c
+1\}}\nonumber\\
&&\times [-3(\ell +1)(2 \ell +1)(-k \eta)^{-\frac{1}{2}}
J_{2\ell + \frac{1}{2}}+ 2 (2\ell +1)(-k \eta)^
{\frac{1}{2}}J_{2\ell -\frac{1}{2}}-(-k \eta)^{\frac{3}{2}}
J_{2 \ell - \frac{3}{2}}]
\nonumber\\
&&- \frac{\sqrt{2\pi}}{3}\frac{C_2 H}{k}
\frac{H r_c}{2 H r_c -1} \sum_{\ell = 0}^{\infty}
\frac{(-1)^{\ell} (2 \ell + \frac{3}{2})}{(2 \ell+1)\{2 \ell  Hr_c
+1\}}\nonumber\\
&&\times [-3 (\ell +1)(2 \ell +3) (-k\eta)^{-\frac{1}{2}}
J_{2\ell + \frac{3}{2}} +2 (2\ell + 2)(-k \eta)^{\frac{1}{2}}
J_{2 \ell + \frac{1}{2}}- (-k \eta)^{\frac{3}{2}} J_{2 \ell
-\frac{1}{2}}],
\label{xi_plus}
\end{eqnarray}
where for $Hr_c \neq 1$
\begin{eqnarray}
\xi_{(m^2 = 2 H^2)} 
&=&- \frac{C_1 H}{6k} \frac{H r_c}{(2H r_c-1)(Hr_c -1)}
\bigl[3(-k\eta)^{-1}
\sin (-k \eta) \nonumber\\
&&+\alpha(-k \eta) \{3(-k\eta)^{-1} \sin (-k \eta) -
3 \cos (-k \eta) - (-k \eta) \sin (-k \eta)\}
\nonumber\\
&&
+\beta(-k \eta)\{3(-k\eta)^{-1} \cos(-k\eta)
+3 \sin(-k\eta) -(-k\eta) \cos (-k\eta)\}\bigr],
\end{eqnarray}
and for $Hr_c =1$
\begin{eqnarray}
\xi_{(m^2 = 2 H^2)} 
&=& \frac{C_1 H}{6k}\Bigl[\tilde{\alpha}(-k\eta)
\{3(-k\eta)^{-1}\sin(-k\eta) -3\cos(-k\eta)
-(-k\eta)\sin(-k\eta)\}\nonumber\\
&&+\tilde{\beta}(-k\eta)\{3(-k\eta)^{-1}\cos(-k\eta) + 3\sin(-k\eta)
-(-k\eta)\cos(-k\eta)\}\nonumber\\
&&
+\frac{3}{2}\alpha(-k\eta)(-k\eta)^{-1}\sin(-k\eta)
+\frac{3}{2}\beta(-k\eta)(-k\eta)^{-1}\cos(-k\eta)\nonumber\\
&&
+3(-k\eta)^{-1}\sin(-k\eta) \Bigr],
\end{eqnarray}

There are homogeneous solutions that satisfy ${\cal{G}}=0$. These are 
given by
\begin{equation}
\Omega  \propto (-k\eta)^{-3/2} Z_{\nu} (- k \eta),
\end{equation}
where $Z_{\nu}$ is a linear combination of Bessel functions and 
$\nu$ is determined by the spectrum of massive modes. 
On small scales, regardless of the value of $\nu$, the 
solution for the brane bending from the homogeneous solutions 
are given by
\begin{equation}
\xi= d_1 (-k \eta) \sin(-k \eta) + d_2 (-k \eta) \cos(-k \eta),
\label{homo}
\end{equation}
where $d_1$ and $d_2$ are arbitrary constants.

\subsection{Perturbation of Brans-Dicke scalar}
Now we examine the behavior of 
$\delta \varphi$ in the BD theory in the presence of the scalar field 
$\phi$, whose dynamics is completely the same as the brane-world 
model. Since we are interested in the behavior of
$\delta \varphi$ on sub-horizon scale, we take the small scale limit.
Then, from Eqs.~(\ref{junc_2_0i}) and 
(\ref{g_junc_cond}), $\phi' \delta \phi'$
can be evaluated as 
\begin{eqnarray}
\phi' \delta \phi' = -\frac{kH}{2r_c \kappa_4^2}
\left[C_1 \cos (-k\eta) + C_2 \sin (-k\eta)\right],
\label{dphi_deltadphi}
\end{eqnarray}
where in order to obtain Eq.~(\ref{dphi_deltadphi}),
we used the fact that $(\ddot{\phi}/\dot{\phi})\ll H \ll k/a$
in the de Sitter background and subhorizon scale.

This term serves as a source term in the equation of
motion of the perturbed Brans-Dicke scalar field
(\ref{field_eq_BD_varphi}) which can be simplified
in the de Sitter background as
\begin{eqnarray}
\delta\varphi'' + \frac{\eta}{2} \delta \varphi' -
\nabla^2 \delta\varphi = -2\zeta^2 \phi' \delta\phi',
\label{field_eq_BD_dS_varphi}
\end{eqnarray}
where $\zeta$ is a constant related with the BD parameter $\omega$
as $\zeta^{2}=(6+4\omega)^{-1}$.
Then by solving Eq.~(\ref{field_eq_BD_dS_varphi})
and keeping leading terms,
we obtain
\begin{eqnarray}
\delta \varphi &=& \tilde{d}_1 (-k \eta)
\sin (-k \eta) + \tilde{d}_2 (- k \eta) \cos (- k \eta) \nonumber\\
&&+\frac{1}{2}\left[ 
 -{\bar{C}}_2 (-k \eta)
\sin (-k \eta) + 
{\bar{C}}_1 (-k \eta)
\cos (-k \eta)\right] Si (-2 k \eta)\nonumber\\
&&+\frac{1}{2} \left[-{\bar{C}}_1 (-k \eta)
\sin (-k \eta) -{\bar{C}}_2 (-k \eta) 
\cos (-k\eta)\right] Ci (-2 k \eta)
\nonumber\\
&& +\frac{1}{2}\left[ {\bar{C}}_1 (-k \eta) 
\ln (-k \eta) \sin (-k \eta) -{\bar{C}}_2 (-k \eta) 
\ln (-k \eta)\cos (-k \eta)\right],
\label{deltaphi_BD}
\end{eqnarray}
where $\tilde{d}_1$ and $\tilde{d}_2$ are integration constants
related with the homogeneous solutions and 
${\bar{C}}_1$ and ${\bar{C}}_2$
are given by
\begin{eqnarray}
{\bar{C}}_1 = \frac{C_1 H \zeta^2} {r_c \kappa_4^2 k},
\;\;\;\;\;\;
{\bar{C}}_2 = \frac{C_2 H \zeta^2} {r_c \kappa_4^2 k}.
\label{def_barC}
\end{eqnarray}
Note that the homogeneous solutions behave 
in the same ways as the brane bending mode coming from 
homogeneous solutions for $\Omega$ (\ref{homo}). These homogeneous solutions are
fixed by appropriate boundary conditions.

\subsection{Numerical results}
Here, we identify the value of the Brans-Dicke 
parameter $\omega$. For this purpose, we evaluate 
Eqs.~(\ref{xi_minus}), (\ref{xi_plus}) and 
(\ref{deltaphi_BD}) numerically.
In practice, we must approximate the infinite sum to proceed 
the calculations. In our previous paper \cite{KMW},
we checked  that we can approximate the infinite summation
by introducing a cut-off $\ell_c$ into the summation
because Bessel function satisfies $J_{\nu}(z) \to 0$ for $\nu > z$.
Therefore,  as long as we start from a finite time $-k \eta_i$,
we can approximate the infinite ladder of the modes 
by introducing sufficiently large cut-off $\ell_c$ and the result 
is insensitive to $\ell_c$.

Here, we compare the part of the numerical solutions
including $C_1$ of Eqs.~(\ref{xi_plus}) with
(\ref{deltaphi_BD}). The same result applies to the 
solution including $C_2$.
For the numerical calculation, we construct the following
dimensionless quantities;
\begin{eqnarray}
\tilde{\xi}_1 = \frac{k}{2C_1 H}\frac{1}{Hr_c}\xi_1
= -\frac{1}{2}\frac{k}{C_1 H^2}(A_{(b)1}+
{\cal{R}}_{(b)1} ),
\nonumber\\
\delta \tilde{\varphi}_1 = \frac{\kappa^2 k} {2 C_1 H}
\delta \varphi_1= 
 -\frac{1}{2}\frac{k}{C_1 H^2}(A_{{\rm BD}1}+
{\cal{R}}_{{\rm BD}1}),
\end{eqnarray}
where the subscript $1$ denotes that they correspond to the
part including $C_1$. 

\begin{figure}[h]
\centerline{
\includegraphics[width=10cm]{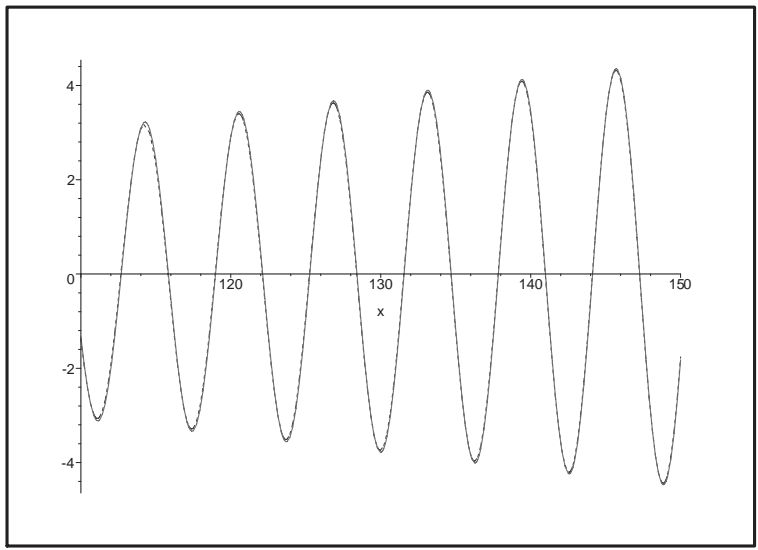}}
\caption{$\tilde{\xi}_1$ (solid line) and 
$\delta \tilde{\varphi}_1$ (dotted line)
as  functions of $x = -k\eta$. The
 Brans-Dicke parameter is $3Hr_c$.
Here we take $Hr_c = 2$ .
The dotted line completely coincides with the
solid line on small scales (large $x$).
The deviation become obvious around $x < 120$.}
\label{fig3}
\end{figure}

Figure 3 shows $\tilde{\xi}_1$ (solid line) and 
$\delta \tilde{\varphi}_1$ (dotted line)
as  functions of $-k\eta$
in the scale smaller than the Hubble scale ($-k\eta =1$).
The Brans-Dicke parameter is chosen as 
$3Hr_c$. Here we take $Hr_c = 2$ and choose the amplitudes
of the homogeneous solutions for $\delta \tilde{\varphi}_1$,
$\tilde{d}_1$ and $\tilde{d}_2$ so that it agrees 
to $\tilde{\xi}_1$ with $d_1=d_2=0$ at $-k\eta = 200$ .
We see that the behavior of perturbations on scales
much smaller than the Hubble scale is well described
by the Brans-Dicke theory with  
$\omega =3 Hr_c$.  
It can also be seen that, at least for $-k\eta < 120$,
the deviation of $\tilde{\xi}_1$ from 
$\delta \tilde{\varphi}_1$ become large and the solution is 
no longer well described by the Brans-Dicke theory with 
$\omega = 3Hr_c$.
Figure 4 also shows  $\tilde{\xi}_1$ (solid line) and 
$\delta \tilde{\varphi}_1$ (dotted line) with $\omega=3Hr_c$
as functions of $-k\eta$ for $H r_c = 0.1$. 
We also choose the homogeneous solutions so that both 
coincide at $-k\eta = 200$ in the same way.
Again we see the solution is well described by the BD theory 
with $\omega =3 Hr_c$, but the deviation of $\tilde{\xi}_1\delta$ from 
$ \tilde{\varphi}_1$ become large for $-k\eta < 180$.
This is natural because, for $Hr_c = 0.1$, $r_c$ becomes 
smaller than the horizon scale (corresponding to $-k\eta = 10$).

\begin{figure}[h]
\centerline{
\includegraphics[width=10cm]{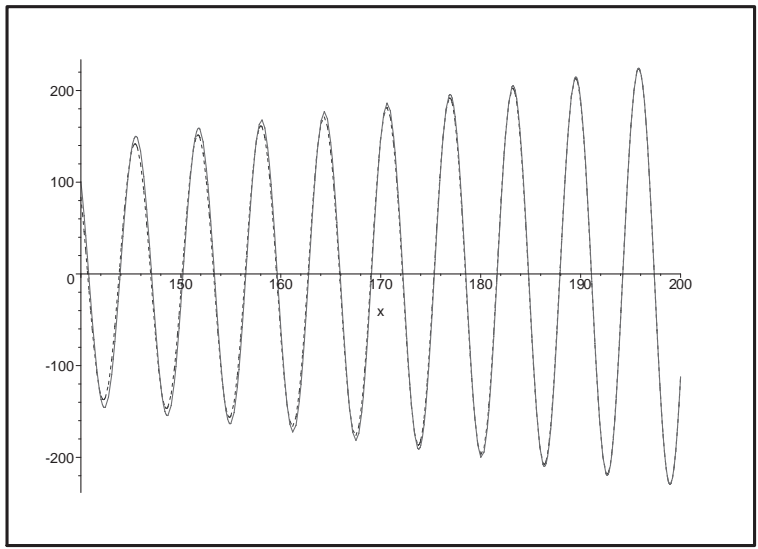}}
\caption{$\tilde{\xi}_1$ (solid line) and 
$\delta \tilde{\varphi}_1$ (dotted line)
as  functions of $x = -k\eta$. The
 Brans-Dicke parameter is $3Hr_c$.
Here we take $Hr_c = 0.1$ .
The dotted line completely coincides with the
solid line on small scales (large $x$).
The deviation become obvious around $x \sim 180$.}
\label{fig4}
\end{figure}

These results are consistent with Ref. \cite{Lue_Starkman, KM}.
In Ref. \cite{Lue_Starkman}, it is shown that the linear 
growth of density contrast in non-relativistic spherical collapse 
of dust is described the Brans-Dicke theory with $\omega = 3Hr_c$ 
for $1/H \gg r$ when $Hr_c$ is greater than $1$.

Even though we do not show the results explicitly here,
we can confirm that, in the $+$ branch, similar
results can be obtained that the perturbations 
can be described by the Brans-Dicke theory with 
$\omega = -3H r_c$ on small scale.
This is also consistent with \cite{Lue_Starkman,KM}.
In the Brans-Dicke theory, the Brans-Dicke scalar
$\varphi$ becomes a ghost if the Brans-Dicke parameter 
satisfies $\omega + 3/2 <0$. 
This result implies that the brane bending
$\xi$ is a ghost for $Hr_c >1$ in $+$ branch.

\section{Curvature perturbation on the brane on large scales}
\subsection{Solutions on Large scales}
In this section, we study the behavior of perturbations
on large scales $(-k\eta) \ll 1$. In this limit, 
the solution for $\Omega$ is dominated by the $m^2 = 2 H^2$ mode .
From Eq.~(\ref{junc_2_0i}) in the limit  
$(-k\eta) \ll 1$, $C_1$ is related with the scalar field
perturbation as
\begin{eqnarray}
\label{scalar_junction}
\kappa^2 \delta \phi = \frac{C_1}{k} \frac{H^2}{\dot{\phi}},
\end{eqnarray}
in both branches.

In the following, we express the curvature perturbation
on the brane as the perturbation of the scalar field 
in this limit in both branches.

\vspace{3mm}
{\bf $-$ branch ($\epsilon = -1$)}
\vspace{2mm}

In this limit, the  $m^2 = 2 H^2$ mode of 
${\cal{R}}_{(b)(m^2 = 2H^2)}$ is,
\begin{eqnarray}
{\cal{R}}_{(b)(m^2 = 2H^2)}  = -  \frac{C_1}{2k} 
\frac{H^2}{(2 r_c H + 1)},
\end{eqnarray}
where we have used the fact that 
$J_{1/2}(-k\eta) \to (-k\eta)^{1/2}\sqrt{2/\pi}$ 
as $(-k\eta) \to 0$.
Then we can relate the perturbation of the scalar field
and the curvature perturbation as
\begin{eqnarray}
{\cal{R}}_{(b)} = -\kappa_{4,-{\rm eff}}^2 \frac{\dot{\phi}}{2H}  \delta \phi, \quad \kappa_{4,-{\rm eff}}^2 = 
\kappa_4^2 \frac{2H r_c}{(2 H r_c +1)},
\end{eqnarray}
which 
is the same as the standard four-dimensional result
(see Eq.~(\ref{curv_pert_4DGR_dS_LS_fluc})), 
except for the overall normalization of the
effective gravitational constant.
If we take the limit $r_c H \to \infty$,
$\kappa_{4,-{\rm eff}}$ coincides with 
$\kappa_{4}$ and we recover four-dimensional results.
We can also show that ${\cal R}_{(b)} = - A_{(b)}$.

\vspace{3mm}
{\bf $+$ branch ($\epsilon = 1$)}
\vspace{2mm}

In this limit,  for $Hr_c \neq 1$, the $m^2 = 2 H^2$ mode of 
${\cal{R}}_{(b)(m^2 = 2H^2)}$ becomes
\begin{eqnarray}
{\cal{R}}_{(b)(m^2 = 2H^2)} = 
-\frac{C_1}{2k}\frac{H^2}{(2Hr_c-1)}
\frac{Hr_c}{(Hr_c-1)},
\end{eqnarray}
where we have used the fact that 
\begin{eqnarray}
\alpha(-k\eta) &\to& 2 \ln (-k\eta)
\label{asym_alpha},\\
\beta(-k\eta) &\to& 2k\eta\;\;\;{\rm as} 
\;\;\;-k\eta \to 0.
\label{asym_beta}
\end{eqnarray}

Strictly speaking, the asymptotic value of $\alpha(-k\eta)$
and $\beta(-k\eta)$ contain an arbitrary integration
constants (see Eqs.~(\ref{alpha_def}) and (\ref{beta_def}).
It was shown in Ref.~\cite{Gen:2000nu, YK} that this degree of freedom corresponds
to a radion and dark radiation. In the following, we neglect this contribution.

Then,  for $Hr_c \neq 1$, 
we can relate the perturbation of the scalar field
and the curvature perturbation as
\begin{eqnarray}
{\cal{R}}_{(b)} = - \kappa_{4,+{\rm eff}}^2 \frac{H r_c}{(H r_c
-1)} \frac{\dot{\phi}}{2H}\delta \phi,
\quad \kappa_{4,+{\rm eff}}^2 =\kappa_4^2  \frac{2 Hr_c}{(2 Hr_c -1)}.
\end{eqnarray}
On the other hand we get
\begin{equation}
A_{(b)} = \kappa_{4,+{\rm eff}}^2 
\frac{\dot{\phi}}{2H}\delta \phi.
\end{equation}
Again, if we take the limit $H r_c \to \infty$,
we recover the four-dimensional results.

\subsection{Self-accelerating background}
Similarly, for $Hr_c=1$,
${\cal{R}}_{(b)(m^2 = 2H^2)}$  becomes
\begin{eqnarray}
{\cal{R}}_{(b)(m^2 = 2H^2)} = 
\frac{C_1 H^2}{2k} (-2 +\ln(-k\eta)),
\end{eqnarray}
where we have used the fact that
\begin{eqnarray}
\tilde{\alpha}(-k\eta) &\to& (\ln(-k\eta))^2
\label{asym_tilde_alpha},\\
\tilde{\beta}(-k\eta) &\to& - 2k\eta
-2\ln(-k\eta)(-k\eta)\;\;\;{\rm as} 
\;\;\;-k\eta \to 0,
\label{asym_tilde_beta}
\end{eqnarray}
as well as Eqs.~(\ref{asym_alpha}) and (\ref{asym_beta}).
In terms of the perturbation of the scalar field,
this is given by
\begin{eqnarray}
{\cal{R}}_{(b)} = \kappa_4^2\frac{\dot{\phi}}{H}
(-2 + \ln (-k\eta)) \delta \phi.
\label{Rb_ito_dphi_delphi_large_hrc1}
\end{eqnarray}
We also get
\begin{equation}
A_{(b)}=0.
\end{equation}
The solution for ${\cal R}_{(b)}$ is diverging on large scales limit 
$(-k \eta) \to 0$. In the next section,
we study whether this means the physical instability on the brane
or not.

\subsection{Effective theory on a brane}
In order to see the origin of the difference between $+$ branch 
and $-$ branch, it is 
useful to consider effective covariant gravitational field equations on the brane
derived by projecting the five-dimensional Einstein
equations and applying the Israel junction conditions
with reflection symmetry at the brane
\cite{SMS,MMT},
\begin{eqnarray}
G_{\mu \nu} =(16 \pi G r_c)^2 \Pi_{\mu\nu}
-E_{\mu\nu},
\label{Einstein_brane}
\end{eqnarray}
where
\begin{eqnarray}
\tilde{T}_{\mu\nu} &=& T_{\mu \nu} - (8 \pi G)^{-1}
G_{\mu\nu}
\label{def_tilde_t_munu},
\\
\Pi_{\mu\nu} &=& -\frac{1}{4}\tilde{T}_{\mu\alpha}
\tilde{T}_\nu^{\;\;\alpha} + \frac{1}{12}
\tilde{T}_\alpha^{\;\;\alpha} \tilde{T}_{\mu\nu}
+\frac{1}{24}\left[3 \tilde{T}_{\alpha\beta}
 \tilde{T}^{\alpha\beta} - 
( \tilde{T}_\alpha^{\;\;\alpha})^2\right]
g_{\mu\nu},
\label{def_pi_munu}
\end{eqnarray}
and $E_{\mu\nu}$ is the trace-free projection of the
five-dimensional Weyl tensor.
In the background, we take $E_{\mu \nu}=0$ by assuming a 
Minkowski bulk. However, perturbations of $E_{\mu \nu}$
are not necessarily vanishing. For perturbations, 
the effective equations are 
\begin{equation} 
\left(1- \frac{1}{2 Hr_c} \right) 
\left\{   
-6H(\dot{{\cal{R}}}_{(b)} - H A_{(b)})- \frac{2 k^2}{a^2} 
{\cal{R}}_{(b)} \right\}
 = - \kappa_4^2 \delta \rho
+ \frac{\kappa_4^2}{2 H r_c} \delta \rho_{E},
\end{equation}
\begin{equation}
2 \left(1- \frac{1}{2 Hr_c} \right) (\dot{{\cal{R}}}_{(b)}-
H A_{(b)}) = \kappa_4^2 
a \delta q - \frac{\kappa_4^2}{2 H r_c} a \delta q_E,
\end{equation}
\begin{equation}
-\frac{1}{a^2} \left(   
1- \frac{1}{2H r_c} \right)
(A_{(b)}+{\cal{R}}_{(b)})
= \kappa_4^2 \delta \pi - 
\frac{\kappa_4^2 \delta \pi_{E}}{2 H r_c}.
\label{anisotropy}
\end{equation}
Substituting the solution for ${\cal R}_{(b)}$ and $A_{(b)}$, we 
find the perturbations of Weyl tensor vanish in $-$ branch.
In $+$ branch, $\delta \rho_E=\delta q_E=0$, but there is 
a non-trivial anisotropic stress;
\begin{eqnarray}
\delta \pi_{E} &=& 
-\frac{1}{a^2} \frac{r_c \dot{\phi}}{(Hr_c-1)} \delta \phi,\quad
Hr_c \neq 1, \\ 
\delta \pi_{E} &=&
\frac{1}{a^2} \frac{\dot{\phi}}{H} \left(\log(-k \eta) -2 \right) \delta \phi,
\quad Hr_c=1.
\end{eqnarray}
Since $a \propto(-k \eta)^{-1}$, even though $\ln(-k\eta)$
diverges as $-k \eta \to 0$, the anisotropic stress from 
perturbations of $E_{\mu\nu}$
does not diverge in the limit $-k\eta \to 0$. 

Since gauge invariant perturbations of Weyl tensor 
do not diverge, the divergence of the curvature perturbation in 
longitudinal gauge Eq.~(\ref{Rb_ito_dphi_delphi_large_hrc1})
does not imply the instability of the spacetime
itself and the divergence is caused simply by a bad choice of the gauge.

In order to show this, let us consider a four-dimensional
gauge transformation,
\begin{equation}
\eta \to \eta - \epsilon^{\eta},\;\;\;\;\;\;
x^i \to x^i - \epsilon^{,i}.
\end{equation}

By choosing $\partial_\eta \epsilon = \epsilon^{\eta}$,
the metric perturbations are transformed as
\begin{eqnarray}
ds^2 = a^2 \{-(1+2\tilde{A}_{(b)})d\eta^2 +
[(1+2\tilde{{\cal R}}_{(b)}) \delta_{ij}+
\tilde{E}_{(b)\:,ij}] dx^i dx^j\},
\end{eqnarray}
where
\begin{eqnarray}
\tilde{A}_{(b)} &=& A_{(b)} -\partial_\eta \epsilon^{\eta}
-(\partial_\eta \ln a)\epsilon^{\eta},\\
\tilde{{\cal R}}_{(b)} &=& {\cal R}_{(b)} - 
(\partial_\eta \ln a)\epsilon^{\eta}, \\
\tilde{E}_{(b)} &=& -\epsilon.
\end{eqnarray}

For $Hr_c \neq 1$, we can eliminate the singular 
part in ${\cal R}_{(b)}$ in the limit $Hr_c \to 1$
by choosing
\begin{equation}
\epsilon^{\eta} = -\kappa_{4,+{\rm eff}}^2 \frac{1}{(Hr_c-1)}
\frac{\dot{\phi}}{2kH} (- k \eta) \delta \phi.
\end{equation}
Then the resultant metric perturbations are 
given by
\begin{eqnarray}
\tilde{{\cal R}}_{(b)} &=& - \tilde{A}_{(b)} 
= -\kappa_{4,+{\rm eff}}^2 \frac{\dot{\phi}}{2 H} \delta \phi, \\
\partial_{\eta} \tilde{E}_{(b)} &=& \kappa_{4,+{\rm eff}}^2 (-k \eta)
\frac{1}{(Hr_c-1)} \frac{\dot{\phi}}{2 k H} \delta \phi.
\end{eqnarray}
For $Hr_c=1$, we can also eliminate the growing part of
${\cal R}_{(b)}$ by choosing 
\begin{eqnarray}
\epsilon^\eta =\frac{\kappa_4^2  \dot{\phi}}{kH}
\delta \phi (-k\eta) \left( \ln(-k\eta) -1 \right).
\end{eqnarray}
Then, the resultant metric perturbations become
\begin{eqnarray}
\tilde{{\cal R}}_{(b)} &=& - \tilde{A}_{(b)} 
= -\kappa_{4}^2 \frac{\dot{\phi}}{H} \delta \phi, \\
\partial_\eta \tilde{E}_{(b)} &=& -\frac{\kappa_4^2 \dot{\phi}}{k H}
(-k\eta) \left(\ln(-k\eta)-1 \right) \delta \phi .
\end{eqnarray}
At late times $-k\eta \to 0$, $E_{(b)}$ behaves as
\begin{eqnarray}
\tilde{E}_{(b)} \propto (-k\eta)^2 \ln(-k\eta) \to 0.
\end{eqnarray}
Thus we can find a gauge where all perturbations remain small
and so there is no gravitational instability.

\section{Conclusion}
In this paper, we studied inflaton perturbations 
confined to a de Sitter brane with induced 
gravity in a five-dimensional Minkowski spacetime.

For a vacuum brane, the spin-0 mode appears as a 
discrete bulk mode with $m^2 = 2H^2$ in the $+$
branch, while in the $-$ branch there are no normalizable
solutions for the spin-0 modes. Since there is another
discrete bulk mode (helicity-0 mode of spin-2 perturbation) 
with mass $m^2 = H^2 (3 Hr_c -1) (Hr_c)^{-2}$ in the $+$ branch, 
there is a resonance
between the spin-0 mode and the helicity-0 mode of spin-2
perturbation for $Hr_c =1$. In this paper we introduced 
inflaton perturbations on a brane. 
Then an infinite ladder of discrete modes
with $m^2 = -2(2\ell-1)(\ell+1)H^2$ and 
$m^2 = -2\ell (2\ell +3)H^2$ are excited. 
Since there is a mode with $m^2 = 2H^2$ regardless of the value
of $Hr_c$, in the $+$ branch the resonance inevitably appears.
We obtained the solutions for the curvature perturbation on the brane 
${\cal{R}}_{(b)}$ and studied their behavior.

At high energies, $H r_c \to \infty$, 
we have confirmed that, in both branches, the four-dimensional general 
relativity solutions are recovered from the infinite sum 
of the modes. This results remind us that in the case of massive gravity,
if we introduce a cosmological constant $\Lambda_{(4)}$,
the vDVZ discontinuity disappears if the limits $m \to 0$ is taken for 
fixed $\Lambda_{(4)}$ \cite{Porrati:2000cp, Higuchi,Kogan:2000uy}.
However, this argument holds only for $Hr_c \to \infty$. For a finite $Hr_c$,
if we do $not$ fix $-k\eta$ and take large enough $-k\eta$,
we can no longer perform
the summation of the infinite ladder of 
the modes and the theory deviate from four-dimensional general relativity.
In fact, on small scales, for a fixed $Hr_c$ 
we can see that the junction
condition becomes the same as the gravitational field
equations in Brans-Dicke theory where the brane bending mode
acts as the BD scalar.

Then, in order to identify the BD parameter for the
effective BD theory, we compared the solutions for 
Brans-Dicke scalar field and the brane bending mode.
We have confirmed that, on sufficiently small scales,
gravity is well described by the Brans-Dicke theory
with $\omega = -3Hr_c$ in $+$ branch and 
$\omega = 3Hr_c$ in $-$ branch. 
This is consistent with the previous work 
\cite{Lue_Starkman, KM} and the existence of the ghost 
in $+$ branch is confirmed.  

We also studied the large scales perturbations. 
In $-$ branch, the solutions agree with four-dimensional GR 
with a modified gravitational constant. In 
$+$ branch, there is an additional contribution 
to ${\cal R}_{(b)}$ due to the resonance, which 
diverges on large scales for $Hr_c=1$.
We identified this contribution as the effect of 
the Weyl anisotropic stress and showed that the 
anisotropic stress itself does not diverge.
In fact, we can find a suitable gauge 
where all metric perturbations remain small, so 
the resonance does not lead to a gravitational instability
on the brane.

We make comments on future applications of our results.
In $-$ branch, if $r_c \ll H^{-1}$, small scale
perturbations $a/k \ll r_c$ can be described by the 
four-dimensional Brans-Dicke theory.  If the perturbations
approaches to $r_c$, the gravity becomes five-dimensional and 
we expect significant effects from the coupling to 
five-dimensional metric perturbations \cite{KMW}. Unlike the 
Randall-Sundrum model where the small scales 
perturbations are always coupled to five-dimensional perturbations
and a quantum vacuum state is hard to be specified 
\cite{KMW, K_Mennim_W, YK},
we can specify a vacuum state without ambiguity 
based on the four-dimensional BD theory on sufficiently small
scales. Then we can estimate the effect of 
the coupling to five-dimensional gravity without ambiguity.

Finally we should emphasize that our analysis is 
limited to linear perturbations. It has been shown that we need to 
take into account the non-linearity of the 
brane bending modes before the gravity 
becomes non-linear. Actually, we can recover 
four-dimensional general relativity due to 
this non-linearity  
\cite{Deffayet_vDVZdis,Lue_string,Gruzinov:2001hp,Giannakis:2001jg,
Porrati:2002cp,Park:2003fc,Tanaka_DGP}.
Ref.~\cite{Lue_Starkman} suggested that this non-linear scale 
becomes very small at high energies by the factor $1 - \epsilon 2 Hr_c$,
which is the same factor seen in Eq.~(\ref{bendcurve}). 
The non-linear interactions of the brane bending mode in 
de Sitter spacetime deserve further study. 

\acknowledgements

SM is grateful to the ICG, Portsmouth for their hospitality when this
work was initiated. SM is supported in part
by the Japan Society for Promotion of Science (JSPS)
Research Fellowship. KK is supported by PPARC.

\appendix
\section{Derivation of $\Omega_{m^2 = 2H^2}$ in $+$ branch}

In order to obtain the mode corresponding to $m^2 = 2H^2$,
we must prepare the solution other than $N(y)$, since
the junction conditions cannot be satisfied by this
solution. We find that depending on whether $Hr_c =1$
or not, there are two types of the solutions.

\vspace{3mm}
{\bf $Hr_c \neq 1$}
\vspace{2mm}

First, we consider the case $Hr_c \neq 1$.
In this case, we assume the following form of the solution:
\begin{eqnarray}
\label{omega_m^2_2h^2}
\Omega_{(m^2 = 2H^2)} =(1+Hy) P(t) + (1+Hy) \ln (1+Hy) Q(t).
\end{eqnarray}
By substituting Eq.~(\ref{omega_m^2_2h^2}) into the junction condition
(\ref{g_junc_cond}), we obtain the following relation;
\begin{eqnarray}
\label{a_b_junc_cond}
(1-2H r_c)H Q(t) -r_c(2H^2-m^2)P(t) = C_1 \sqrt{\frac{\pi}{2}}
(-k\eta)^{-\frac{3}{2}} J_{\frac{1}{2}} (-k\eta).
\end{eqnarray}
It is worth noting that in Eq.~(\ref{a_b_junc_cond}), since $P(t)$
is no longer $(-k\eta)^{-3/2} J_{\frac{1}{2}} (-k\eta)$,
$m^2$ is not a constant but a differential operator.
On the other hand, we find that the equation of the motion in the bulk
is satisfied if the following conditions hold:
\begin{eqnarray}
\label{a_bulk_cond}
(2H^2 -m^2) P(t) &=& -H^2 Q(t),\\
\label{b_bulk_cond}
(2H^2 -m^2) Q(t) &=& 0.
\end{eqnarray}

By eliminating $P(t)$ from Eqs.~(\ref{a_b_junc_cond}) and 
(\ref{a_bulk_cond}), we get
\begin{eqnarray}
\label{b_conc}
Q(t) =- C_1 \frac{1}{H(H r_c-1)}
(-k\eta)^{-2} \sin (-k\eta),
\end{eqnarray}
and by substituting Eq.~(\ref{b_conc}) into Eq.~(\ref{a_bulk_cond})
we obtain
\begin{eqnarray}
\label{a_conc}
P(t)=C_1 \frac{1}{2H(H r_c-1)} \frac{1}{(-k\eta)^2}
\{ \alpha(-k\eta) \sin(-k\eta) + \beta(-k\eta) \cos(-k\eta)\},
\end{eqnarray}
where 
\begin{eqnarray}
\label{alpha_def}
\alpha(-k\eta) &=& \int^{-k\eta} d(-k\bar{\eta}) (-k\bar{\eta})^{-2}
\sin (-2k\bar{\eta}),\\
\label{beta_def}
\beta(-k\eta) &=& 
\int^{-k\eta} d(-k\bar{\eta}) (-k\bar{\eta})^{-2}
\{\cos (-2k\bar{\eta})-1\}.
\end{eqnarray}

\vspace{3mm}
{\bf $Hr_c = 1$}
\vspace{2mm}

Next, we consider the case $Hr_c = 1$.
In this case, we assume the following form of the solution:
\begin{eqnarray}
\label{omega_m^2_2h^2_hrc1}
\Omega_{(m^2 = 2H^2)} =(1+Hy) P(t) + (1+Hy) \ln (1+Hy) Q(t)
+ (1+Hy)\left(\ln (1+Hy)\right)^2 S(t).
\end{eqnarray}
By substituting Eq.~(\ref{omega_m^2_2h^2_hrc1})
into the junction condition
(\ref{g_junc_cond}), we can obtain the following relation:
\begin{eqnarray}
\label{a_b_junc_cond_hrc1}
-H Q(t) -\frac{1}{H}(2H^2-m^2)P(t) = C_1 \sqrt{\frac{\pi}{2}}
(-k\eta)^{-\frac{3}{2}} J_{\frac{1}{2}} (-k\eta).
\end{eqnarray}
It is worth noting that in Eq.~(\ref{a_b_junc_cond_hrc1}), 
since $P(t)$
is no longer $(-k\eta)^{-3/2} J_{\frac{1}{2}} (-k\eta)$,
$m^2$ is not a constant but a differential operator.
On the other hand, we find that the equation of the motion 
in the bulk
is satisfied if the following conditions hold:
\begin{eqnarray}
\label{a_bulk_cond_hrc1}
(2H^2 -m^2) P(t) &=& (m^2 - 3H^2)Q(t),\\
\label{b_bulk_cond_hrc1}
(2H^2 -m^2) Q(t) &=& -2 H^2 S(t),\\
\label{c_bulk_cond_hrc1}
(2H^2 -m^2) S(t) &=& 0.
\end{eqnarray}

By eliminating $P(t)$ and $Q(t)$ from 
Eqs.~(\ref{a_bulk_cond_hrc1}), (\ref{b_bulk_cond_hrc1})
and (\ref{c_bulk_cond_hrc1}) we obtain
\begin{eqnarray}
\label{c_conc_hrc1}
S(t) = -\frac{C_1}{2H} 
(-k\eta)^{-2} \sin(-k\eta).
\end{eqnarray}
By substituting (\ref{c_conc_hrc1}) into 
(\ref{b_bulk_cond_hrc1}), we obtain
\begin{eqnarray}
\label{b_conc_hrc1}
Q(t) = \frac{C_1}{2H}\frac{1}{(-k\eta)^2}
\{\alpha(-k\eta) \sin(-k\eta) + \beta(-k\eta)
\cos (-k\eta)\},
\end{eqnarray}
where $\alpha(-k\eta)$ and $\beta(-k\eta)$ are
defined as Eqs.~(\ref{alpha_def}) and (\ref{beta_def}).
Furthermore, by substituting (\ref{c_conc_hrc1})
and (\ref{b_conc_hrc1}) into (\ref{a_bulk_cond_hrc1}),
we get 
\begin{eqnarray}
\label{a_conc_hrc1}
P(t) = -\frac{C_1}{2H} \frac{1}{(-k\eta)^2}
\{\tilde{\alpha}(-k\eta) \sin(-k\eta) 
+ \tilde{\beta}(-k\eta) \cos(-k\eta)\},
\end{eqnarray}
where
\begin{eqnarray}
\label{til_alpha_def}
\tilde{\alpha}(-k\eta) &=& \alpha(-k\eta)+\frac{1}{2}
\int^{-k\eta} d(-k\bar{\eta}) (-k\bar{\eta})^{-2}
[\alpha(-k\bar{\eta}) \sin(-2k\bar{\eta})
+\beta (-k\bar{\eta}) \{\cos(-2k\bar{\eta})+1\}],\\
\label{til_beta_def}
\tilde{\beta}(-k\eta) &=& \beta(-k\eta) +\frac{1}{2}
\int^{-k\eta} d(-k\bar{\eta}) (-k\bar{\eta})^{-2}
[\alpha(-k\bar{\eta}) \{\cos(-2k\bar{\eta})-1\}
-\beta (-k\bar{\eta}) \sin(-2k\bar{\eta}) ].
\end{eqnarray}

\section{Metric perturbations in 4D Brans-Dicke gravity}
In this appendix, we briefly summarize the
results for cosmological perturbations in four-dimensional Brans-Dicke theory
(see \cite{Fujii_Maeda} for a review of BD theory).

In this theory, we start with the following action:
\begin{eqnarray}
{\cal{L}}_{\rm BD} = \sqrt{-g}\left(\varphi R - \omega \frac{1}{\varphi}
g^{\mu \nu} \partial_\mu \varphi \partial_\nu \varphi
+ L_{\rm matter}\right),
\label{action_BD}
\end{eqnarray}
where $\varphi$ is a Brans-Dicke scalar field and $\omega$ is 
a Brans-Dicke
coupling constant.
From Eq.~(\ref{action_BD}) we can derive the field equations,
\begin{eqnarray}
2\varphi G_{\mu\nu} &=& T_{\mu\nu} + T_{\mu\nu}^{\varphi}
-2(g_{\mu\nu} \Box - \nabla_\mu \nabla_\nu) \varphi,\label{g_field_eq_BD}\\
\Box \varphi &=& \zeta^2 T,\label{s_field_eq_BD}\\
\nabla_\mu T^{\mu\nu} &=& 0,\label{cons_eq_BD}
\end{eqnarray}
where $\zeta$ is a constant 
related with $\omega$ as
\begin{eqnarray}
\zeta^{-2} = 6 + 4 \omega.
\end{eqnarray}

For perturbations, the metric is taken as 
\begin{eqnarray}
ds^2 = 
-(1+2A) dt^2 + a(t)^2 (1+2 {\cal{R}}) \delta_{ij}dx^i dx^j.
\label{4d_lond_gauge}
\end{eqnarray}
When we consider a canonical scalar field 
with a potential $V(\phi)$ as the matter, 
the field equations reduce to
\begin{eqnarray}
&&6{\cal{H}}^2 \delta \varphi + 4 \varphi_0[-3 {\cal{H}}
({\cal{H}} A -{\cal{R}}') - \nabla^2 {\cal{R}}]
=-{\phi'}^2 A + \phi' \delta \phi' + {V,}_\phi a^2 \delta \phi
-6{\cal{H}} \delta \varphi' + 2\nabla^2 \delta \varphi,
\label{field_eq_BD_00}\\
&&4 \varphi_0 ({\cal{H}} A - {\cal{R}}' ) = \phi' \delta \phi
+ 2 \delta \varphi' - 2 {\cal{H}} \delta \varphi,
\label{field_eq_BD_0i}\\
&&2 \delta \varphi (2{\cal{H}}' + {\cal{H}}^2) -4 \varphi_0
[(2{\cal{H}}' + {\cal{H}}^2)A + {\cal{H}} A' - {\cal{R}}''-
2{\cal{H}} {\cal{R}}' + \frac{1}{2} \nabla^2 (A + {\cal{R}})]
\nonumber\\
&&\;\;\;\;\;\;\;\;\;\;=
{\phi'}^2 A - \phi' \delta \phi' + {V,}_\phi a^2 \delta \phi
-2 \delta \varphi'' -2{\cal{H}} \delta \varphi' + 2 \nabla^2 \delta \varphi,
\label{field_eq_BD_ii}\\
&&\varphi_0 (A+{\cal{R}}) = - \delta \varphi,
\label{field_eq_BD_ij}\\
&&\delta \varphi'' + 2{\cal{H}} \delta \varphi'
-\nabla^2 \delta \varphi =
2 \zeta^2 [{\phi'}A - \phi' \delta \phi + 2 {V,}_\phi a^2 \delta \phi]
\label{field_eq_BD_varphi},
\end{eqnarray}
where the prime here denotes the derivative with respect to
conformal time $\eta$ and ${\cal{H}} = a'/a$

If we require Eqs.~(\ref{field_eq_BD_00}), 
(\ref{field_eq_BD_0i}),
(\ref{field_eq_BD_ii}) and (\ref{field_eq_BD_ij}) reproduces 
the results of GR in the limit of $\delta \varphi \to 0$
and $\omega \to \infty$, the value of 
$\varphi_0$ is set to be $\varphi_0 = (1/2\kappa_4^2)$.
For the scale much smaller than the Hubble radius
and the scalar field evolves slowly in the background,
Eqs.~(\ref{field_eq_BD_00}), (\ref{field_eq_BD_0i}),
(\ref{field_eq_BD_ii}) and (\ref{field_eq_BD_ij}) can be simplified as 
\begin{eqnarray}
-4\varphi_0 \nabla^2 {\cal{R}} &=& \phi' \delta \phi'
+ 2 \nabla^2 \delta \varphi,
\label{field_eq_BD_00_s}\\
-4\varphi_0 {\cal{R}}' &=& \phi' \delta \phi +2 \delta\varphi',
\label{field_eq_BD_0i_s}\\
4\varphi_0  {\cal{R}}'' - 2 \varphi_0 \nabla^2(A+ {\cal{R}})
&=& -\phi' \delta \phi' + 2\nabla^2 \delta \varphi,
\label{field_eq_BD_ii_s}\\
\varphi_0 (A+{\cal{R}}) &=& - \delta \varphi.
\label{field_eq_BD_ij_s}
\end{eqnarray}
Then the equation for curvature perturbation is obtained as
\begin{eqnarray}
{\cal{R}}'' - \nabla^2  {\cal{R}} = \frac{1}{2\varphi_0}
\nabla^2 \delta \varphi.
\label{b_eq_curv_pert_4DBD_dS}
\end{eqnarray}

If we take general relativity limit, we get
\begin{eqnarray}
{\cal{R}}'' -\nabla^2 {\cal{R}} =0,
\label{b_eq_curv_pert_4DGR_dS}
\end{eqnarray}
and its Fourier modes of ${\cal{R}}$ are  expressed as
\begin{eqnarray}
{\cal{R}} = \tilde{C_1} \cos (-k\eta) + \tilde{C_2}
\sin (-k\eta),
\label{curv_pert_4DGR_dS}
\end{eqnarray}
where $\tilde{C_1}$ and $\tilde{C_2}$
are integration constants \cite{Mukhanov}.
The lapse function $A$ is obtained as 
\begin{eqnarray}
A = -{\cal{R}} =  -\tilde{C_1} \cos (-k\eta) - \tilde{C_2}
\sin (-k\eta).
\label{lapse_func_4DGR_dS}
\end{eqnarray}

If we take a large scale limit, that is, $-k\eta \to 0$,
from Eq.~(\ref{lapse_func_4DGR_dS}), we can obtain
\begin{eqnarray}
{\cal{R}} \to \tilde{C_1} = {\rm const.}
\label{curv_pert_BD_dS_LS}
\end{eqnarray}
From Einstein equations, the curvature perturbation
is related with the scalar field fluctuation as
\begin{eqnarray}
{\cal{R}} = -\kappa_4^2 \frac{\dot{\phi}}{2 H} \delta \phi.
\label{curv_pert_4DGR_dS_LS_fluc}
\end{eqnarray}

\end{document}